\documentclass{aa} 
\usepackage{natbib}
\bibpunct{(}{)}{;}{a}{}{,} 
\usepackage{graphicx}
\usepackage{txfonts}
\newcommand{\fermi}{{\em Fermi}\xspace}

\newcommand{\swift}{{\em Swift}\xspace}

\newcommand{\sw}[1]{\texttt{#1}}
\graphicspath{{./}{figures/}}

\begin{document}

   \title{10.4m GTC observations of the nearby VHE-detected GRB 190829A/SN 2019oyw}


 \author{Y.-D. Hu\inst{\ref{inst1},\ref{inst2}} \and A. J. Castro-Tirado\inst{\ref{inst1},\ref{inst3}} \and A. Kumar\inst{\ref{inst4},\ref{inst5}} \and R. Gupta\inst{\ref{inst4},\ref{inst6}} \and A. F. Valeev\inst{\ref{inst61},\ref{inst63}} \and S. B. Pandey\inst{\ref{inst4}} \and D. A. Kann \inst{\ref{inst1}} \and A. Castellón\inst{\ref{inst62}} \and I. Agudo\inst{\ref{inst1}} \and A. Aryan\inst{\ref{inst4},\ref{inst6}} \and M. D. Caballero-García\inst{\ref{inst9}} \and S. 
 Guziy \inst{\ref{inst68},\ref{inst69}} \and A. Martin-Carrillo\inst{\ref{inst8}} \and S. R. Oates\inst{\ref{inst10}} \and E. Pian\inst{\ref{inst11}} \and R. Sánchez-Ramírez\inst{\ref{inst12}} \and V. V. Sokolov\inst{\ref{inst61}} \and B.-B. Zhang\inst{\ref{inst14},\ref{inst15}} }
 
 \institute{ 
 Instituto de Astrof\'isica de Andaluc\'ia (IAA-CSIC), P.O. Box 03004, E-18080 Granada, Spain \label{inst1} \\ \email{huyoudong072@hotmail.com}
 \and Universidad de Granada, Facultad de Ciencias Campus Fuentenueva s/n E-18071 Granada, Spain \label{inst2}
 \and Departamento de Ingenier\'ia de Sistemas y Autom\'atica, Escuela de Ingenier\'ias, Universidad de M\'alaga, C\/. Dr. Ortiz Ramos s\/n, 29071 M\'alaga, Spain \label{inst3} 
 \and Aryabhatta Research Institute of Observational Sciences (ARIES), Manora Peak, Nainital - 263002, India \label{inst4}
 \and School of Studies in Physics and Astrophysics, Pandit Ravishankar Shukla University, Chattisgarh 492 010, India \label{inst5}
 \and Department of Physics, Deen Dayal Upadhyaya Gorakhpur University, Gorakhpur 273009, India \label{inst6}
 \and Special Astrophysical Observatory of Russian Academy of Sciences, Nizhniy Arkhyz, Russia \label{inst61}
 \and Crimean Astrophysical Observatory, Russian Academy of Sciences, Nauchnyi, 298409 Russia  \label{inst63}
 \and Departamento de Álgebra, Geometría y Topología, Facultad de Ciencias, Universidad de Málaga, Málaga, Campus de Teatinos, E-29071 sn, Málaga, Spain \label{inst62}
 \and Astronomical Institute, Academy of Sciences of the Czech Republic, Bo\v{c}n\'{\i}~II 1401, CZ-141\,00~Prague, Czech Republic. \label{inst9}
 \and Mykolaiv National University, Astronomical Observatory, Mykolaiv, Ukraine \label{inst68}
 \and Research Institute, Mykolaiv Astronomical Observatory, Mykolaiv, Ukraine \label{inst69}
 \and School of Physics, O’Brien Centre for Science North, University College Dublin, Belfield, Dublin 4, Ireland \label{inst8}
 \and School of Physics and Astronomy, University of Birmingham, B15 2TT, UK \label{inst10}
 \and INAF, Astrophysics and Space Science Observatory, via P. Gobetti 101, 40129 Bologna, Italy \label{inst11}
 \and INAF, Istituto di Astrofisica e Planetologia Spaziali, via Fosso del Cavaliere 100, I-00133 Rome, Italy \label{inst12}
 \and School of Astronomy and Space Science, Nanjing University, Nanjing 210093, China \label{inst14} 
 \and Key Laboratory of Modern Astronomy and Astrophysics (Nanjing University), Ministry of Education, China \label{inst15} }

   \date{Received ; 2020/2020 }

 
  \abstract{}
   {GRB 190829A ($z = 0.0785$), detected by \fermi and \swift with two emission episodes separated by a quiescent gap of $\sim$ 40 s, was also observed by the H.E.S.S. telescopes at Very-High Energy (VHE). We present the 10.4m GTC observations of the afterglow of GRB 190829A and underlying supernova and compare it against a similar GRB 180728A and discuss the implications on underlying physical mechanisms producing these two GRBs.}
   {We present multi-band photometric data along with spectroscopic follow-up observations taken with the 10.4m GTC telescope. Together with the data from the prompt emission, the 10.4m GTC data are used to understand the emission mechanisms and possible progenitor.} 
   {A detailed analysis of multi-band data of the afterglow demands cooling frequency to pass between the optical and X-ray bands at early epochs and dominant with underlying SN 2019oyw later on.}
   {Prompt emission temporal properties of GRB 190829A and GRB 180728A are similar, however the two pulses seem different in the spectral domain. We found that the supernova (SN) 2019oyw associated with GRB 190829A, powered by Ni decay, is of Type Ic-BL and that the spectroscopic/photometric properties of this SN is consistent with those observed for SN 1998bw but evolved comparatively early. }


   \maketitle
%

\section{Introduction}

Multi-wavelength observations of nearby (redshift z $\leq$ 0.2) long-duration Gamma-ray bursts (GRBs) and their association with 
Type Ic supernovae with broad lines (Type Ic-BL SNe) have revolutionized our understanding in the explosion mechanisms and environments of 
massive stars across the electromagnetic spectrum \citep{1993ApJ...405..273W, 2012grb..book..169H}. Some of these nearby GRBs also belong to the class of low-intermediate luminosity GRBs and ultra long-duration GRBs, outliers, which have revealed crucial observational evidence to distinguish between potential powering mechanisms and progenitors \citep{Georgy2009,Dessart2017}. As underlying supernova features are faint and diluted by their host galaxies, 8-10m 
class optical-NIR facilities play a vital role to extract information \citep{2013JApA...34..157P, 2014A&A...568A..19C}. So far, there are only handful of nearby GRBs, having associated Ic-BL SNe: GRB 980425/SN 1998bw \citep[$z=0.00867$;][]{1999A&AS..138..465G}, GRB 030329/SN 2003dh \citep[$z=0.16867$;][]{2003ApJ...591L..17S}, GRB 031203/SN 2003lw \citep[$z=0.10536$;][]{2004ApJ...609L...5M}, GRB 060218/SN 2006aj \citep[$z=0.03342$;][]{2006ApJ...643L..99M}, GRB 100316D/SN 2010bh \citep[$z=0.0592$;][]{2010arXiv1004.2262C}, GRB 130702A/SN 2013dx \citep[$z=0.145$;][]{2015A&A...577A.116D}, GRB 171205A/SN 2017iuk \citep[$z=0.0368$;][]{2019Natur.565..324I} and GRB 180728A/SN 2018fip  
\citep[$z=0.117$;][]{2018GCN.23142....1I}.

GRB 190829A belongs to a rare class whose prompt emission light curves have double episodes: the first being a fainter, harder pulse (a precursor) and after a considerable delay a second significantly brighter and softer main pulse. This two pulse event provides a unique opportunity to probe deeper into the nature of the central engine \citep{2014ApJ...789..145H, 2020ApJ...898...42C, 2020arXiv200311252F} of this VHE detected burst \citep{2019GCN.25566....1D}. The proximity of this burst provided the opportunity to discover the underlying SN \citep{2019GCN.25623....1P,2019GCN.25652....1L,2019GCN.25664....1T, 2019GCN.25651....1B,Dichiara2020}, giving the opportunity to address the question as to whether such a two-episodic prompt emission nature has possible connection towards proposed progenitor models \citep{2006ARA&A..44..507W, 2007AIPC..906...69D}. Another notable example is the nearby GRB 180728A ($z=0.117$) associated with SN 2018fip  \citep{2018GCN.23181....1S,2018GCN.23142....1I,2019ApJ...874...39W} exhibiting remarkable similarity with GRB 190829A 
in terms of two episodic prompt emission with a significant temporal gap at {\it Swift}/BAT and {\it Fermi}/GBM frequencies. This prompted us to perform a joint prompt emission analysis of these two nearby events using a sophisticated tool called the Multi-Mission Maximum Likelihood framework (\sw{3ML}\footnote{https://github.com/threeML/threeML}) to explore their spectral properties systematically.

VHE photons from GRB 190829A were detected by the High Energy Stereoscopic System \citep[H.E.S.S.;][]{2019GCN.25566....1D} making this burst the nearest to be seen at these high frequencies. Many ground based telescopes searched for the counterparts soon after the $\it{Swift}$ and $\it{Fermi}$ gamma-ray detection and follow-up observations at other wavelengths were reported. Using the 10.4m Gran Telescopio CANARIAS (GTC, Canary Islands, Spain) optical-NIR observations, a redshift $z$ = 0.0785 $\pm$ 0.005 was reported \citep{2019GCN.25565....1V}, thus triggering larger facilities to follow-up the event and later reporting the re-brightening of the underlying transient AT 2019oyw/SN 2019oyw \citep{2019TNSTR1657....1L, 2019GCN.25664....1T, 2019GCN.25682....1V, 2019GCN.25677....1D}. The late time afterglow observations  at radio frequencies were also reported by \cite{2019GCN.25627....1C}, \cite{2019GCN.25676....1L}, and \cite{2020MNRAS.496.3326R}.

In this paper, we present the analysis of the prompt emission properties of GRB 190829A and GRB 180728A. We also discuss our late time photometric and spectroscopic observations of GRB 190829A/SN 2019oyw using the 10.4m GTC telescope and their comparison with other well-studied similar events. This paper is organized as follows. In \S~\ref{prompt observations and data analyisis}, we present the prompt observations and data analysis of GRB 190829A and GRB 180728A. Then, we focus on the optical spectroscopic observations of GRB 190829A and their analysis in \S~\ref{spectroscopic}. In \S~\ref{Photometric}, we present the optical photometric observations of GRB 190829A. Finally, we discuss and conclude in \S~\ref{results}. All the uncertainties are quoted at 1-$\sigma$ level throughout this paper, unless mentioned otherwise. We assume the Hubble constant $\rm H_{0}$ = 70 km $\rm s^{-1}$ $\rm Mpc^{-1}$ and density parameters $\rm \Omega_{\Lambda}= 0.73$ and $\rm \Omega_m= 0.27$. 


\section{Prompt emission properties: GRB 190829A and GRB 180728A}
\label{prompt observations and data analyisis}
\begin{figure}
 \centering
  \includegraphics[scale=0.38]{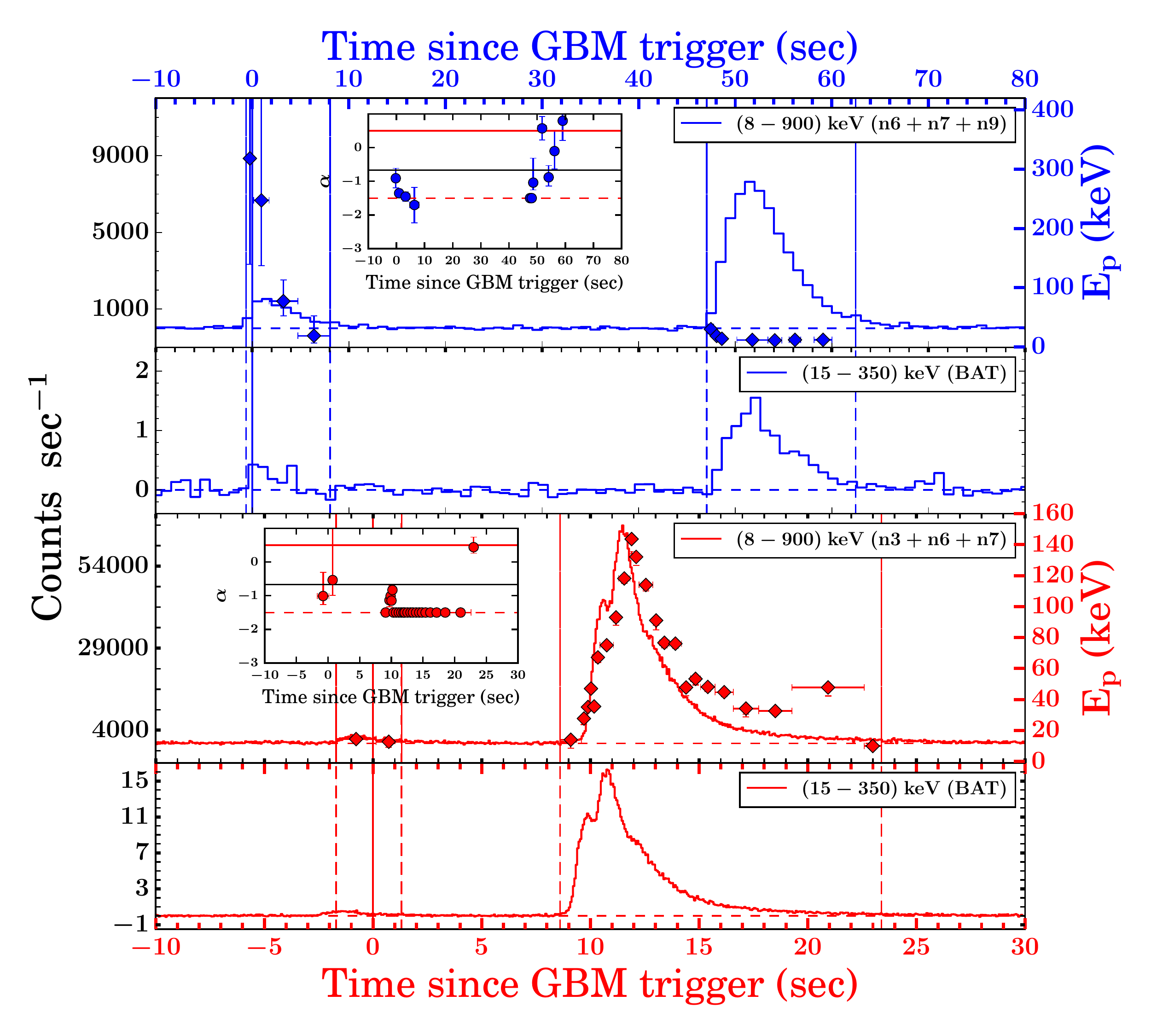}
  \caption{{\bf Prompt emission lightcurves of GRB 190829A (blue) and GRB 180728A (red):} The two upper panels show the gamma-ray lightcurves of GRB 190829A (1 s bins) whereas the two bottom panels show the gamma-ray lightcurves of GRB 180728A (64 ms bins) for the given energy channels. Solid blue and red vertical lines represent {\em Fermi}/trigger times whereas red and blue dashed vertical lines encompass the time interval considered for the joint time-averaged spectral analysis for each episode of the two bursts. The peak energy evolution of the first episode of GRB 190829A shows a hard to soft trend whereas the second episode is more disordered. The behaviour of this second episode is contrary to the $\rm E_{\rm p}$ behaviour observed for GRB 180728A which tracks intensity. The insets for GRB 190829A (blue) and GRB 180728A (red) show the evolution of the low-energy spectral index ($\alpha$) with the red-dashed and the black-solid lines representing the synchrotron fast cooling spectral index ($-3/2$) and the line of death of synchrotron emission ($-2/3$), respectively. Interestingly, in the case of GRB 190829A $\alpha$
  seems to overshoot the synchrotron limits in later bins whereas in the case of GRB 180728A the evolution of $\alpha$ remains within the synchrotron limit.
  }
  \label{promptlc}
\end{figure}

\begin{figure*}
\centering
  \includegraphics[scale=0.38]{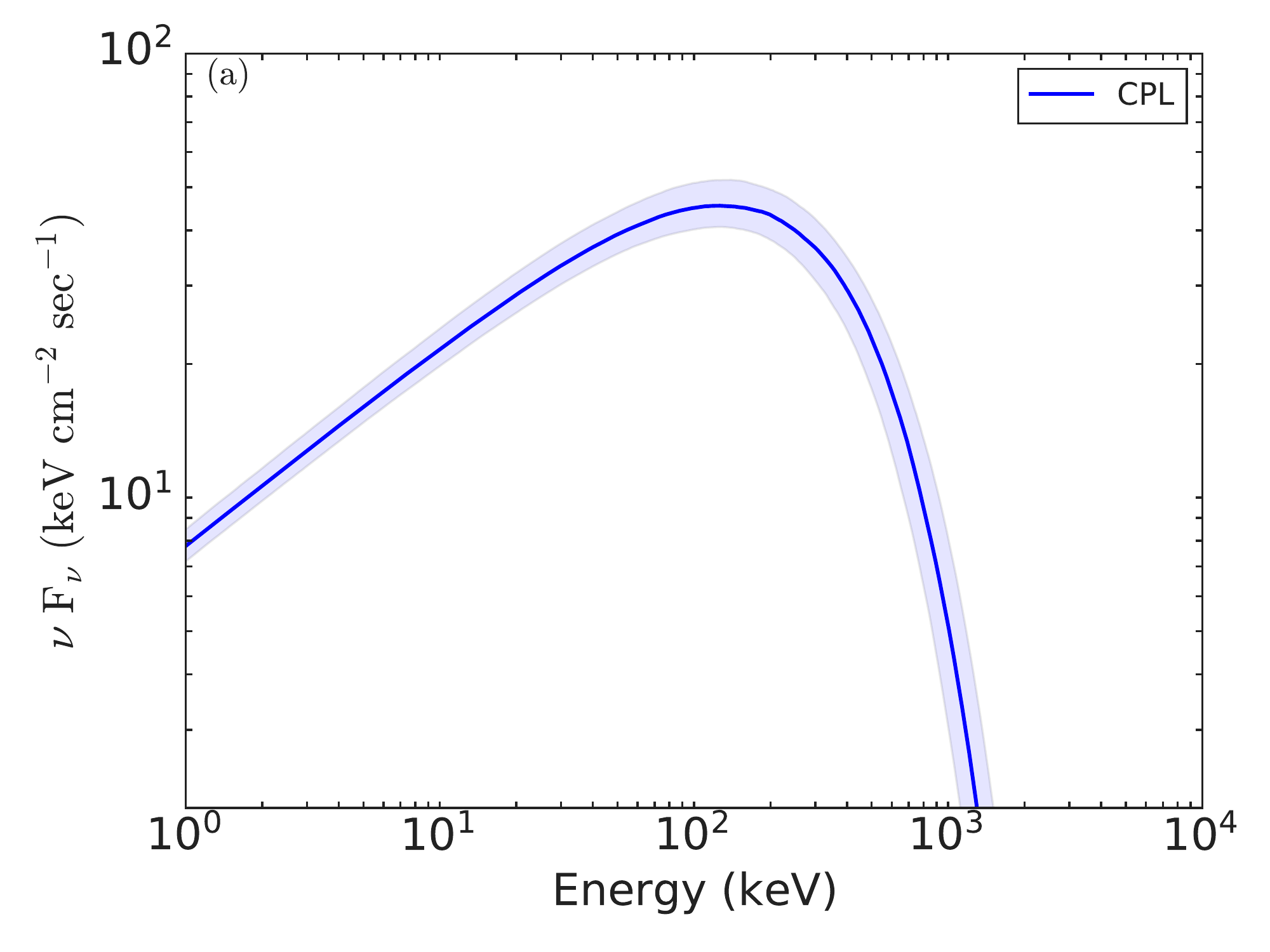}
 \includegraphics[scale=0.38]{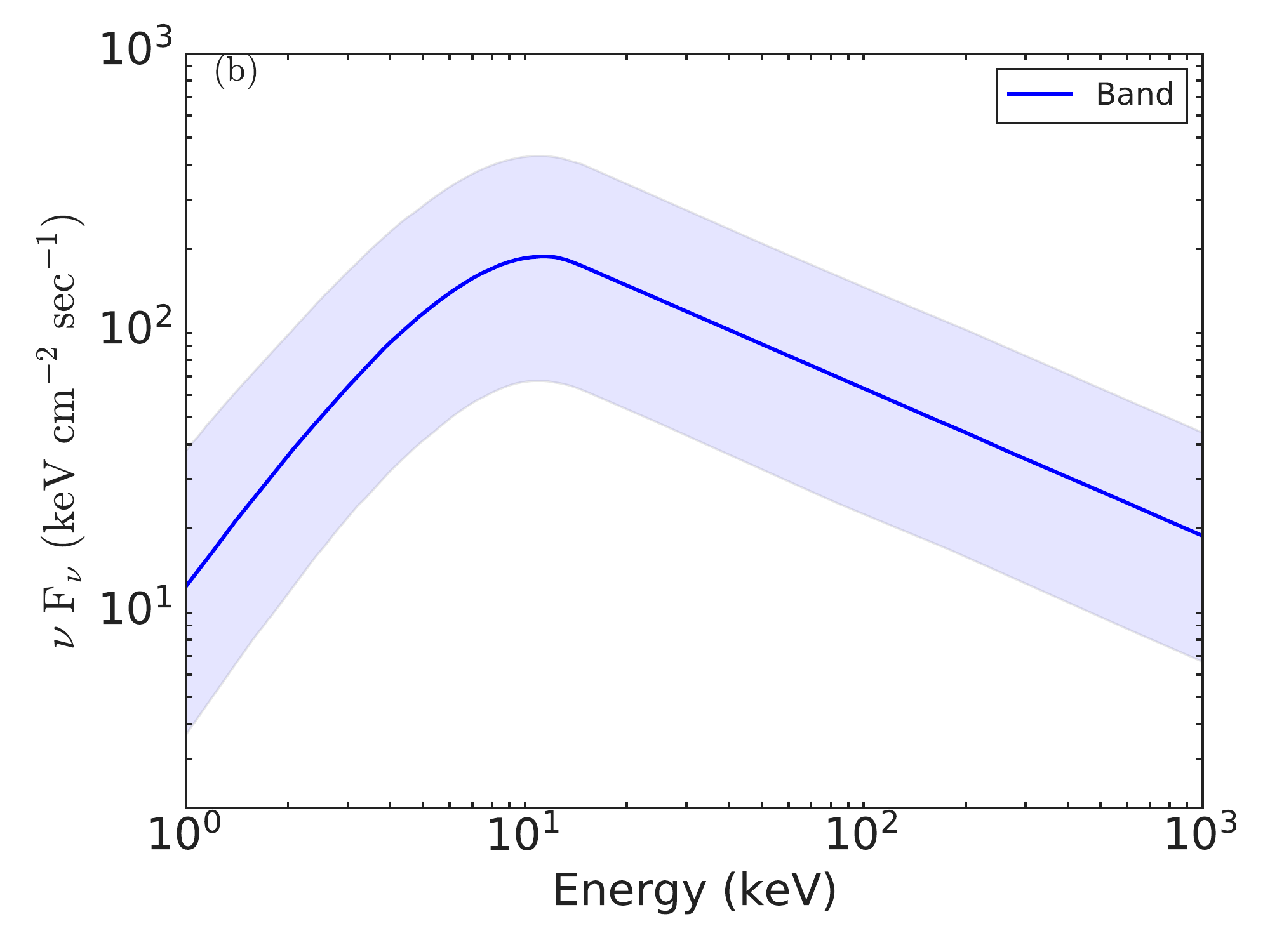}
 \includegraphics[scale=0.38]{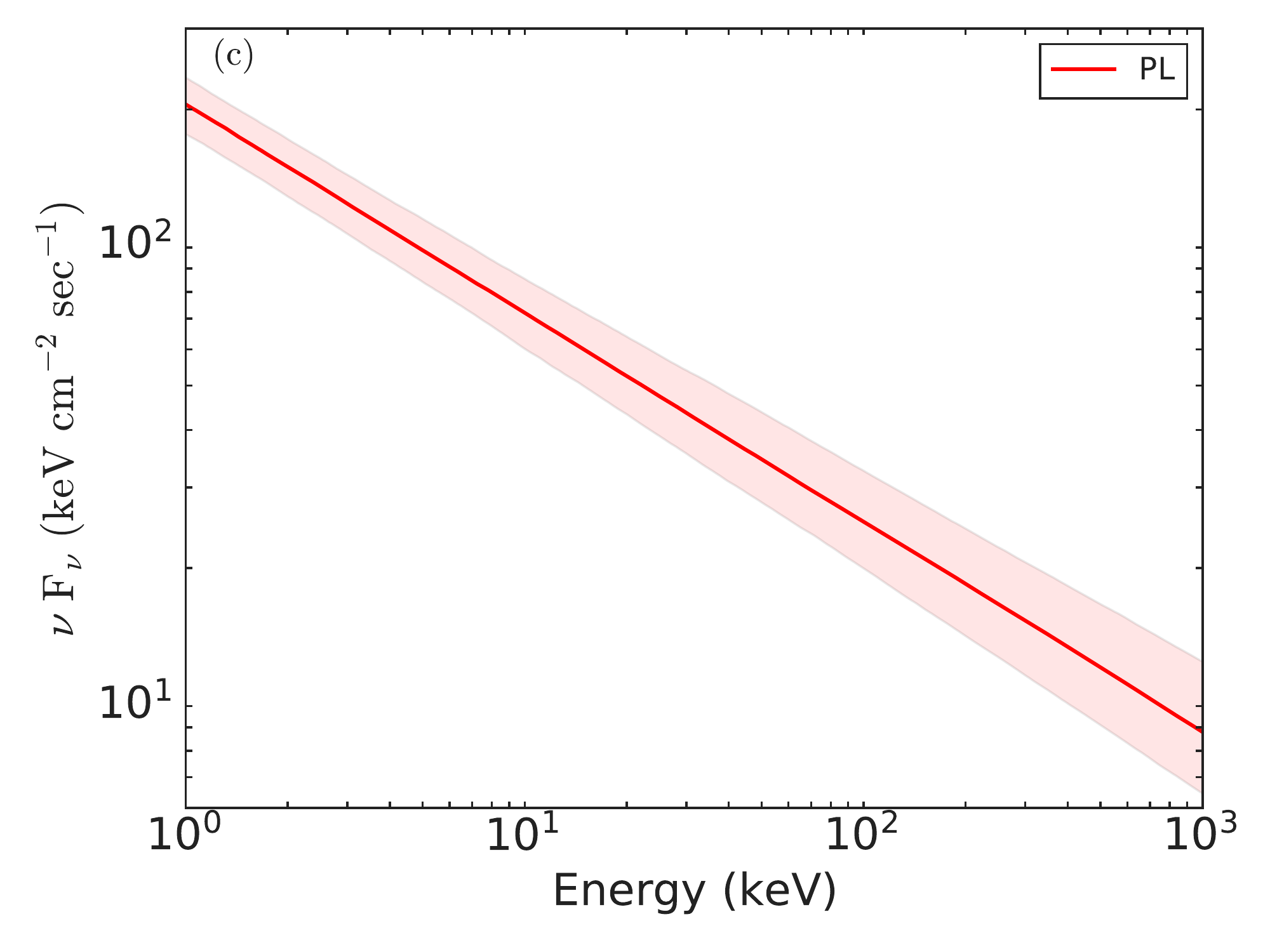}
  \includegraphics[scale=0.38]{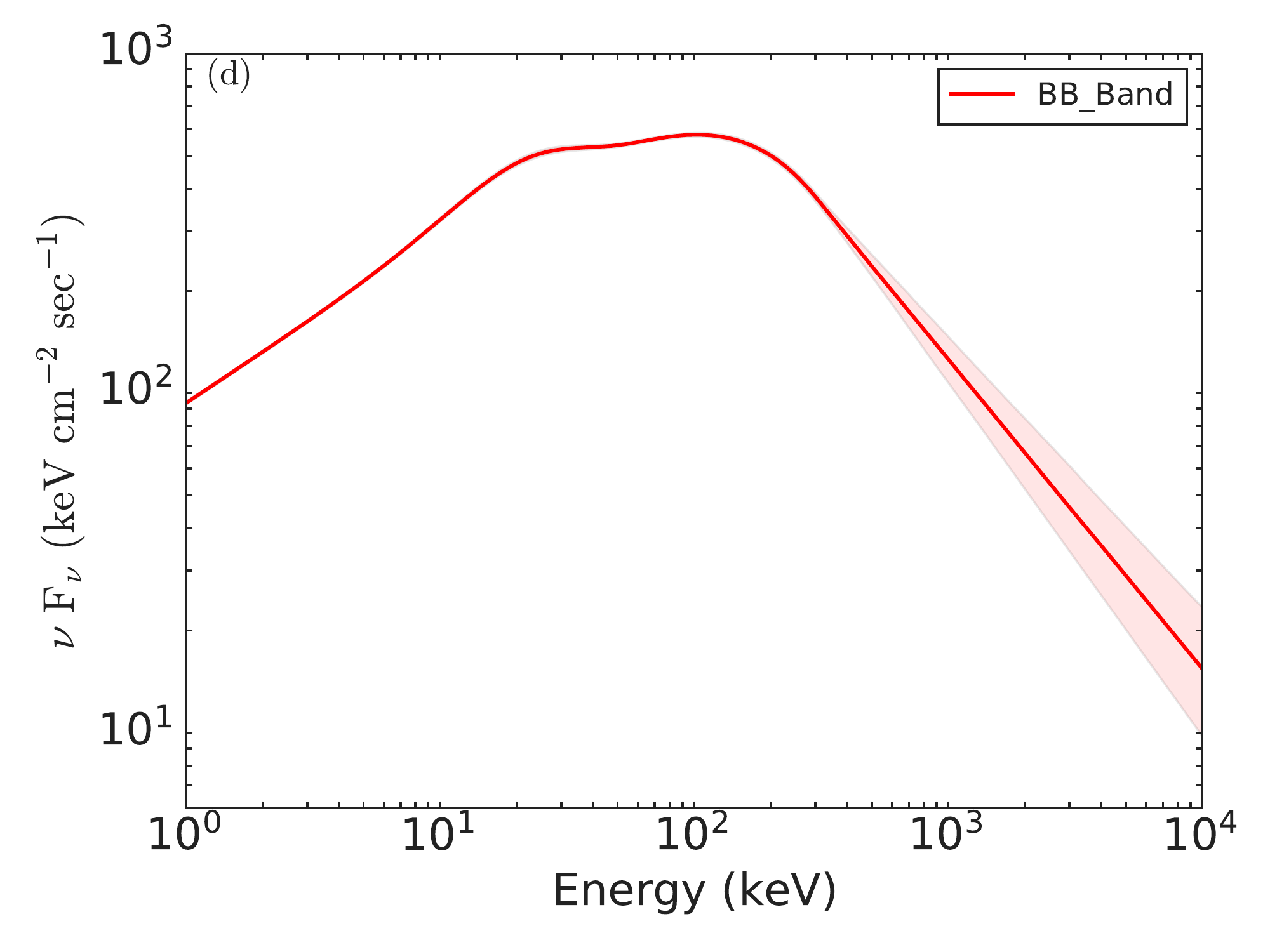}
\caption{The time-integrated best fit energy spectrum of GRB 190829A and GRB 180728A in model space using joint spectral analysis of different combination of detectors (see Table \ref{tab:promptModel}). (a) First pulse of GRB 190829A modelled with a \sw{Cutoff power-law} model. (b) The second pulse of GRB 190829A is best described with a \sw{Band} function. (c) First pulse of GRB 180728A modelled with a \sw{Power-law} model. (d) The second pulse of GRB 180728A described with a combination of \sw{Band} and \sw {BB} functions.}
\label{TAS_fig}
\end{figure*}

\begin{table*}
\caption{Comparison between different models used for episode-wise time-averaged joint spectral analysis of \fermi-GBM and \swift/BAT data of GRB 180728A and GRB 190829A.  In the model column, the best fit model is mentioned with star. PL, CPL, and BB correspond to the Power-law, Cutoff power-law and Blackbody models, respectively.}
\begin{center}
\begin{tabular}{|l|l|l|c|c|c|}

\hline\hline
 &&\textbf{GRB 180728A}&&&\\
\hline\hline
\textbf{Episode}    & \textbf{Time (s)} & \textbf{Model} & \textbf{Log(Likelihood)} & \textbf{AIC$^{a}$} & \textbf{BIC$^{b}$} \\
\hline
 1             & -1.70 - 1.31      & { PL*}       &       1475.81         &  2961.76     &     2981.78 \\
           &                   & CPL            &          1475.81        &  2963.82    &  2987.81  \\
(n3+n6+n7+BAT)              &                   & Band           &   1471.44                 &  2957.15     &   2985.11    \\
                    &                   & PL+BB          &     1475.81              &  2965.89     &  2993.85     \\
                    &                   & CPL+BB         &  1487.62              & 2985.39      & 3005.41      \\
                    &                   & Band+BB        &   1471.24            &   2960.92    &   2996.77    \\
\hline
 2             & 8.61 - 23.39     & PL             &  4358.70                &   8729.56    &  8755.10    \\
(n3+n6+n7+b1+BAT)         &                   & CPL            &  3395.31                  &   6804.84   &  6834.60    \\
                    &                   & Band           &       3472.00            & 6960.27      & 6994.25      \\
                    &                   & PL+BB          &     3877.83             &   7771.93   &  7805.91    \\
                    &                   & CPL+BB         &    3347.10            &   6712.54    &  6750.74    \\
                    &                   & Band+BB* &       3273.44          & 6567.31      &     6609.71 \\ 
\hline\hline
 &&\textbf{GRB 190829A}&&&\\
\hline\hline
 1             & -0.64 - 8.06      & PL      &       2268.45       &    4549.07 &  4574.61   \\
           &                   & CPL*            &            2261.82      &   4537.84   & 4567.62  \\
(n6+n7+n9+b1+BAT)              &                   & Band           &   2262.13                 &  4540.54     &   4574.54     \\
                    &                   & PL+BB          &   2264.48         & 4545.24      & 4579.24     \\
                    &                   & CPL+BB         &    2261.43           & 4541.21   &  4579.43     \\
                    &                   & Band+BB        &    unconstrained         &  unconstrained   &  unconstrained   \\
\hline
 2             & 47.04 - 62.46     & PL             & 2182.14                & 4374.42      &  4394.47    \\
(n6+n7+n9+BAT)         &                   & CPL            & 2175.54                   &  4363.29    &  4387.31    \\
                    &                   & Band*          &           2147.17       &  4308.61    &    4336.60   \\
                    &                   & PL+BB          &  2150.16                 & 4314.60    & 4342.59     \\
                    &                   & CPL+BB         & 2148.33                & 4313.02    & 4344.97  \\
                    &                   & Band+BB  &       2142.71      &   4303.85   &  4339.75    \\ 

\hline\hline

\end{tabular}

\end{center}
\footnotesize{$^a$ Akaike Information Criterion, $^b$ Bayesian Information Criterion.}

\label{tab:promptModel}

\end{table*}
The {\it Fermi} satellite first triggered on GRB 190829A on 29 Aug, 2019 at 19:55:53 UT \citep[T$_0$,][]{2019GCN.25551....1F}. After 51 s, the Burst Alert Telescope (BAT) on-board {\it Swift} also triggered and located this event at 19:56:44 UT with a duration of T$_{90}$  \footnote{T$_{90}$ is burst duration defined as the time interval over which 5\% to 95\% of the counts are accumulated.} = 58.2 $\pm$ 8.9 s~\citep{2019GCN.25552....1D}. Both Fermi/GBM and Swift/BAT  reported the temporal behaviour of this GRB as having a double peaked structure. The X-ray telescope (XRT) on-board {\it Swift} began observing the field 97.3\,s after the BAT trigger and found a bright, fading uncatalogued X-ray source and continued to monitor until four months after the trigger. The astrometrically corrected X-ray position is RA(J2000)=$02^{h}$ $58^{m}$ $10.57^{s}$, Dec(J2000)=$-08^{\circ}$ $57'30.1''$ with an uncertainty of $1.8''$~\citep[90\% confidence radius;][]{2019GCN.25567....1E}.

We obtained the data of the Gamma-ray Burst Monitor (GBM) on-board the \fermi satellite from the GBM archive\footnote{\url{ https://heasarc.gsfc.nasa.gov/W3Browse/fermi/fermigbrst.html}} and analyzed it using the Multi-Mission Maximum Likelihood framework (\sw{3ML}) software. We studied the temporal and spectral prompt emission properties of GRB 190829A and GRB 180728A using data from the three sodium iodide (NaI) detectors with bright detections and the brightest-detected bismuth germanate (BGO) detector (if available) along with \swift/BAT data. These two nearby GRBs have similar temporal behaviour consisting of two episodes, a weak precursor and a main burst, separated by a quiescent gap. This rarely observed temporal behaviour prompted us to compare them together. To perform the time-averaged and time-resolved spectral analysis, we reduced the time-tagged event (TTE) mode data of GBM using the \sw{gtburst}\footnote{\url{https://fermi.gsfc.nasa.gov/ssc/data/analysis/scitools/gtburst.html}} software as they have high time precision in all 128 energy channels. We retrieved the \swift/BAT light curve and spectrum following the standard procedure\footnote{\url{https://www.swift.ac.uk/analysis/bat/index.php}}, fitted with the \sw{Band} function \citep{1993ApJ...413..281B} and various other models (\sw{Black-body}, \sw{Cutoff Power-law}, and \sw{Power-law} or their combinations) based upon the model fit, residuals of the data, and their parameters. The results based on the analysis described above are presented in Figs. \ref{promptlc}, and \ref{TAS_fig}, and the values are tabulated in Table \ref{tab:promptModel}. The episodes are identified with the Bayesian Blocks method~\citep{1998ApJ...504..405S}.

For GRB 190829A, the time-averaged spectrum of the precursor (first episode) is best described as a power-law with an exponential high-energy 
cutoff function having a photon index of $-1.56^{+0.07}_{-0.08}$, and a cutoff energy corresponding to the peak energy, {E$_{\rm p}$} = 123.51$^{+56.14}_{-31.61}$ keV.  On the contrary, the main episode is best fit by a \sw{Band} function  with {E$_{\rm p}$} = 11.23$^{+0.30}_{-0.32}$ $\rm keV$, low-energy spectral index ($\rm \alpha$) = -0.23 $^{+0.26}_{-0.24}$ and high-energy spectral index ($\rm \beta$) = -2.53$^{+0.01}_{-0.01}$, 
consistent with \cite{2019GCN.25575....1L}. However, in the case of GRB 180728A, the precursor episode is best described by a power-law with photon index equal to -2.45$^{+0.04}_{-0.05}$ whereas the  main episode is best described with \sw{band}+ \sw{black-body} component with {E$_{\rm p}$} = 102.70$^{+2.12}_{-2.00}$ $\rm keV$, $\rm \alpha$ = -1.50 $^{+0.01}_{-0.01}$ , $\rm \beta$ = -2.91$^{+0.12}_{-0.12}$ and the temperature (kT) = 5.61$^{+0.09}_{-0.10}$ keV.

\section{10.4m GTC spectroscopic observations of GRB 190829A/SN 2019oyw}
\label{spectroscopic}
\begin{figure}
\centering
\includegraphics[width=16cm,angle=0,scale=.5]{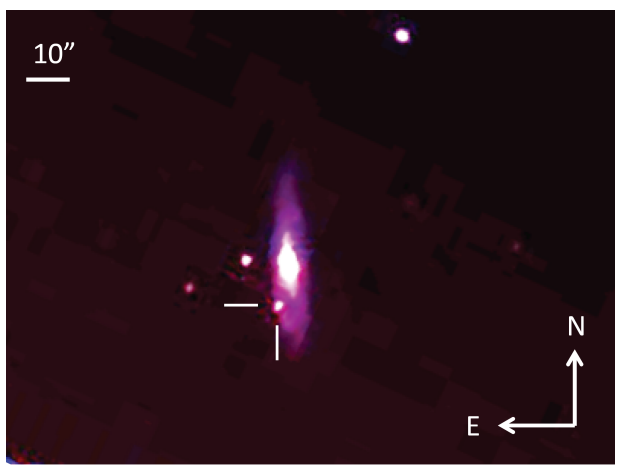}
\caption{The gri-band false colour image of the field of GRB 190829A taken with the 10.4m GTC on Aug 29, 2019. Two perpendicular lines indicate the afterglow position. The associated host galaxy, J025810.28-085719.2, is clearly seen. North is up and east to the left.}
\label{GTC_RGBimage}
\vspace{-0.4cm}
\end{figure}

Spectroscopic observations of GRB 190829A were triggered at the 10.4m GTC (+OSIRIS) soon after the burst and a set of spectroscopic observations were acquired (see Table \ref{tab:spectroscopylog}). We obtained optical spectroscopy covering the range 3,700-10,000 \text{\AA} from 0.32 to 4.09 days post-burst (in the rest-frame), in order to monitor the optical evolution of GRB 190829A/SN 2019oyw, see Table~\ref{tab:spectroscopylog}. Follow-up observations were further possible as the underlying SN 2019oyw lying in the outskirts of the host as shown in Fig. \ref{GTC_RGBimage}. In the following sub-sections, the redshift determination and spectral analysis of the underlying supernova along with the comparison to other similar nearby events are described.

\subsection {Redshift determination:}

\begin{figure}
 \centering
  \includegraphics[scale=0.58]{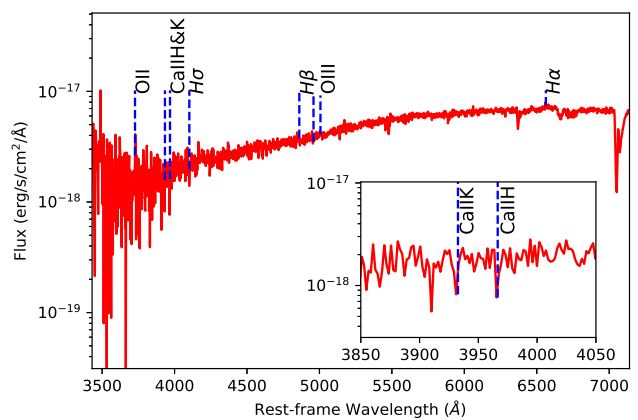}
  \caption{The 10.4m GTC optical spectrum in the range $3,500-7,000$ \AA, which provides the redshift of GRB 190829A. The CaII lines are shown in absorption (see Inset) and the emission lines of the underlying host galaxy at the same redshift.}
  \label{redshiftSpectrum}
\end{figure}

The Ca H \& K absorption lines doublet (3933.664 and 3968.470 \AA{}) were identified in the observed spectrum (see Fig. \ref{redshiftSpectrum}) which allowed us to determine the redshift $z$ = 0.0785 $\pm$ 0.0005 ~\citep{2019GCN.25565....1V}. Emission lines (O\,III, H$\alpha$, H$\beta$) of the underlying galaxy are also visible at the same redshift, thus supporting  the physical association between GRB 190829A and the SDSS galaxy J025810.28-085719.2, as first proposed by \cite{2019GCN.25552....1D}. In order to achieve the maximum spectral resolution at bluer wavelengths we used the R1000B (and especially the R2500U) grisms on 30 Aug, 2019 in order to constrain the redshift.

\subsection{Spectroscopic evolution of GRB 190829A/SN 2019oyw}
\label{opticalsp}

\begin{figure}
 \centering
  \includegraphics[scale=0.9]{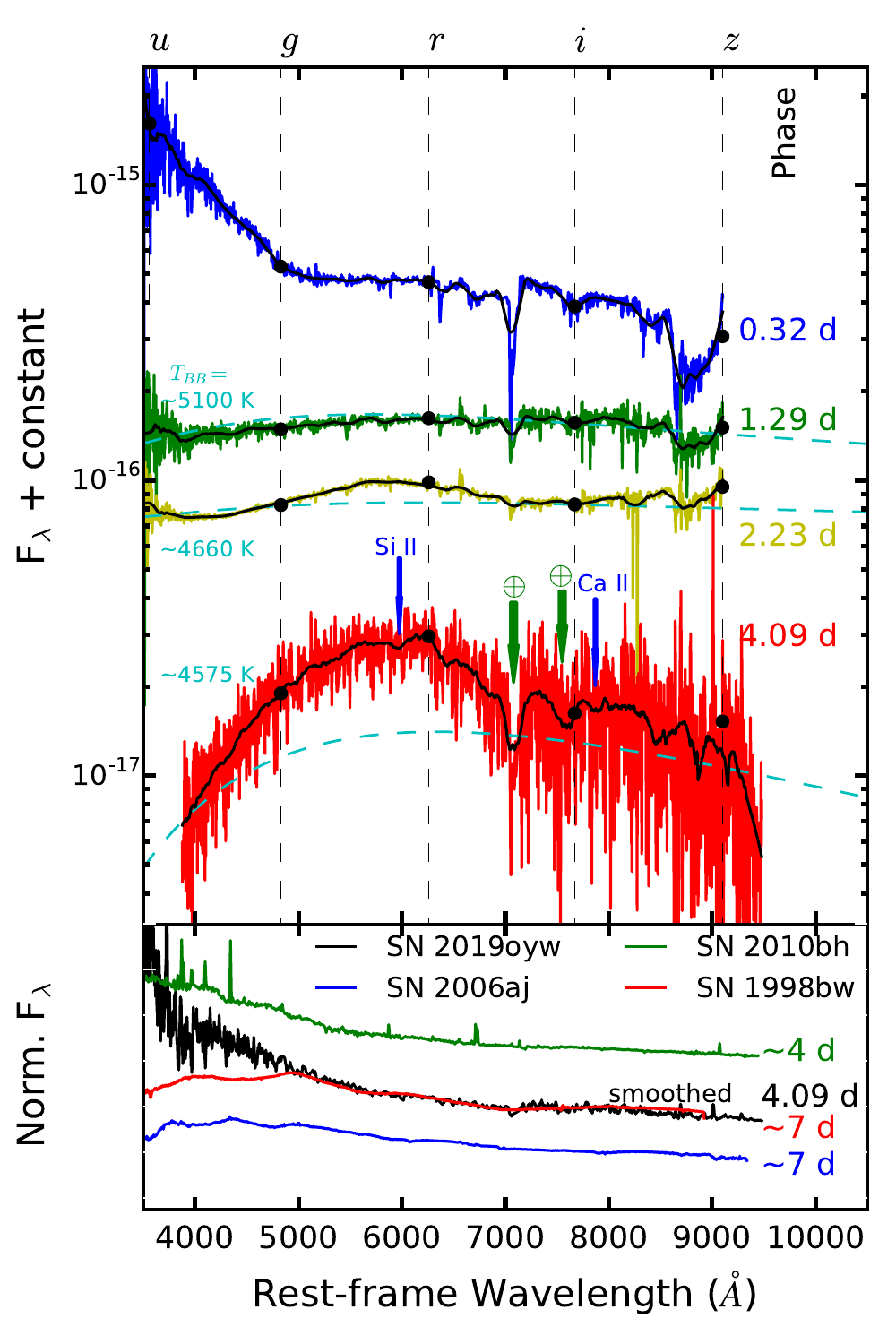}
  \caption{Spectroscopic evolution of GRB 190829A/SN 2019oyw from 0.32 to 4.09 days after the burst. The flux density as a function of the rest-frame wavelength is plotted after correcting spectra for the Galactic and host extinctions.
  All the presented spectra in the upper panel have been corrected for grism losses and scaled to the observed photometric flux density values shown with black circles, wherever possible. For clarity, random offsets in the Y-axis are applied and spectra are also smoothed using the Savitzky–Golay method as described by \cite{Quimby2018}. Un-usual bluer excess in the early spectrum taken at 0.32 day is also noticeable, deviating from the overall power-law nature.}
  \label{spectra}
\end{figure}

The upper panel of Fig. \ref{spectra} shows the spectral evolution. Multiple spectra observed at similar epochs were co-added to improve the signal-to-noise ratio. All the spectra have been de-reddened for the Galactic and host extinction values \citep{Schlafly2011,2020ApJ...898...42C} and also shifted to the rest-frame wavelength. Because of poor signal-to-noise the smoothing of spectra has been done using the Savitzky–Golay method by fitting the second-order polynomial function for each $\lambda$ in the range $\lambda$ $-$ $\lambda$/50 $<$ $\lambda$ $<$ $\lambda$ + $\lambda$/50, as described by \cite{Quimby2018}. All the spectra have been flux calibrated by scaling them to the observed photometric flux density values (shown with black circles in Fig. \ref{spectra}) whenever possible.

As shown in Fig. \ref{spectra}, the first two spectra (at 0.32 and 1.29 days post-burst) are featureless and typical of those expected from GRB afterglows, but a transition from the afterglow (AG) to the underlying supernova (SN) is clearly illustrated with broad lines as emerging features in the later (after $\sim$ 2.23 days) optical spectra of GRB 190829A/SN 2019oyw indicating high velocities already at this stage. The spectrum at 0.32 days shows a power-law behaviour, whereas the spectra at 1.29, 2.23, and 4.09 days appear to  deviate from the power-law and can be constrained with the black-body function implying black-body temperatures ($\rm T_{\rm BB}$) of $\sim$ 5100, $\sim$ 4660, and $\sim$ 4575 K, respectively (shown with cyan color in the upper panel of Fig. \ref{spectra}). The clear atmospheric features are indicated with green arrows in the spectrum taken at 4.09 days. The associated SN 2019oyw spectrum at 4.09 days appears to have Si II ($\lambda$ 6355 \text{\AA}) and Ca II NIR ($\lambda$ 8498 \text{\AA}, 8542 \text{\AA}, and 8662 \text{\AA}) spectral features (blue arrows) at higher velocities, typical of those seen Type Ic-BL SNe.

In the bottom panel of Fig. \ref{spectra}, the spectrum of SN 2019oyw taken at $\sim$ 4.09 days (in black, smoothed) is compared with other GRB/SNe spectra such as SN 1998bw (in red; \citealt{Patat2001}), SN 2006aj (in blue; \citealt{Pian2006}), and SN 2010bh (in green; \citealt{Bufano2012}). For comparison, all the spectra were normalized, de-reddened (Galactic + host) and shifted to the rest-frame wavelength. It is clearly noticeable that the observed broad spectral features for GRB 190829A/SN 2019oyw are similar to those observed in other cases. However, the spectrum of SN 2019oyw taken at $\sim$ 4.09 days shows close resemblance with
the spectrum of SN 1998bw taken at 7 days. This time difference between the spectra of the two SN indicates faster evolution for SN 2019oyw.

\begin{table}
\tiny
 \begin{center}
    \caption{GRB 190829A/SN 2019oyw spectroscopy observation log at the 10.4m GTC.}
    \label{tab:spectroscopylog}
    \addtolength{\tabcolsep}{1pt}
    \begin{tabular}{c c c c c}
\hline
MJD & Phase & Range & Detector & Exp time \\
  & (days)  &(\text{\AA}) &   &(s) \\         
    \hline
   \hline
58725.120411 & 0.293 & 5100-10000 & OSIRIS+R1000R & 600\\
58725.127789 & 0.301 & 3630-7500 & OSIRIS+R1000B & 600\\
58725.180198 & 0.375 & 3440-4610 & OSIRIS+R2500U & 1200X2\\
58725.216805 & 0.393 & 3630-7500 & OSIRIS+R1000B & 1200\\
58726.202041 & 1.387 & 3630-7500 & OSIRIS+R1000B & 900x2\\
58726.223595 & 1.397 & 5100-10000 & OSIRIS+R1000R & 600\\
58727.230284 & 2.405 & 3630-7500 & OSIRIS+R1000B & 900\\
58727.241158 & 2.414 & 5100-10000 & OSIRIS+R1000R & 600\\
58729.228917 & 4.407 & 3630-7500 & OSIRIS+R1000B & 750x2\\
58729.246993 & 4.420 & 5100-10000 & OSIRIS+R1000R & 600\\ 

      \hline 
    \end{tabular}
  \end{center}
\end{table}

\begin{table}
\tiny
 \begin{center}
    \caption{Optical-photometric data of GRB 190829A/SN 2019oyw in SDSS \textit{u}, \textit{g}, \textit{r}, \textit{i}, and \textit{z}-bands obtained using the GTC 10.4m.}
    \label{tab:photodata}
    \addtolength{\tabcolsep}{1pt}
    \begin{tabular}{cccccc}
\hline
MJD & Phase$^{a}$ & Exp time & Filter & mag (AB)$^{b}$  & error \\
     $ $  & (days)  & (s)  &    & (mag)  & (mag)   \\         
    \hline
   \hline
58725.195050 & 0.365 & 120  & \textit{u}  &  21.77 &  0.05  \\
58726.186209 & 1.381 &  120x4 & \textit{u}  &  >22.83 & --   \\
58727.204482 & 2.377 & 120x5 & \textit{u}  &  >23.28 &  -- \\
58728.229954 & 3.400 &  120  & \textit{u}  &  >23.65 &  --  \\
58729.203136 & 4.376 &  120x5 & \textit{u}  &  >23.64 & --  \\
\hline
58725.196904 & 0.367 &  60 & \textit{g}  &  21.08 &  0.05  \\
58726.184475 & 1.382 &  120x2  & \textit{g}  &  23.56 &  0.05  \\
58727.212898 & 2.385 & 120x3 & \textit{g}  &  25.04 &  0.04  \\
58728.231701 & 3.403 &  120x3 & \textit{g}  &  23.95 &  0.06  \\
58729.211539 & 4.383 & 120x3 & \textit{g}  &  24.64 &  0.09  \\
\hline
58725.198025 & 0.368 &  30  & \textit{r}  &  19.68 &  0.01  \\
58726.191965 & 1.364 & 60x2 & \textit{r}  &  21.75 &  0.06  \\
58727.224253 & 2.396 &  180  & \textit{r}  &  22.60 &  0.03  \\
58728.236767 & 3.407 & 120  & \textit{r}  &  22.94 &  0.03  \\
58729.222909 & 4.394 &  60x2  & \textit{r}  &  22.59 &  0.06  \\
\hline
58725.115196 & 0.287 &  10x3 & \textit{i}  &  18.40 &  0.02  \\
58725.176400 & 0.354 & 50  & \textit{i}  &  18.79 &  0.01  \\
58726.183430  & 1.382 &  60x2 & \textit{i}  &  20.73 &  0.05  \\
58727.217984 & 2.388 &  120 & \textit{i}  &  22.34 &  0.06  \\
58728.238499 & 3.409 & 120  & \textit{i}  &  22.30 &  0.02  \\
58729.216606  & 4.387 &  120  & \textit{i}  &  22.15 &  0.08  \\
58732.158872  & 7.331 & 30x3 & \textit{i}  &  22.00 &  0.04  \\
58736.180631 & 11.35 &  30x2 & \textit{i}  &  21.08 &  0.03  \\
58739.131398 & 14.30 & 30x2 & \textit{i}  &  20.76 &  0.09  \\
58754.138870 & 29.31 &  60x2  & \textit{i}  &  21.23 &  0.02  \\
58765.162492 & 40.33 & 60  & \textit{i}  &  22.48 &  0.01  \\
\hline
58725.199912 & 0.370 &  30  & \textit{z}  &  18.17 &  0.02  \\
58726.177119 & 1.347 &  30 & \textit{z}  &  19.93 &  0.00  \\
58727.219749  & 2.390 & 45x3  & \textit{z}  &  20.70 &  0.01  \\
58728.240237 & 3.411 & 45x3  & \textit{z}  &  20.44 &  0.04  \\
58729.218345  & 4.389 &  45x3  & \textit{z}  &  21.31 & 0.05   \\
      \hline 
    \end{tabular}
    \footnotesize{$^a$ Time after the burst, $^b$ Galactic extinction corrected.}
      \label{tab:photometry}
  \end{center}

\end{table}

\begin{figure}
 \centering
  \includegraphics[scale=0.33]{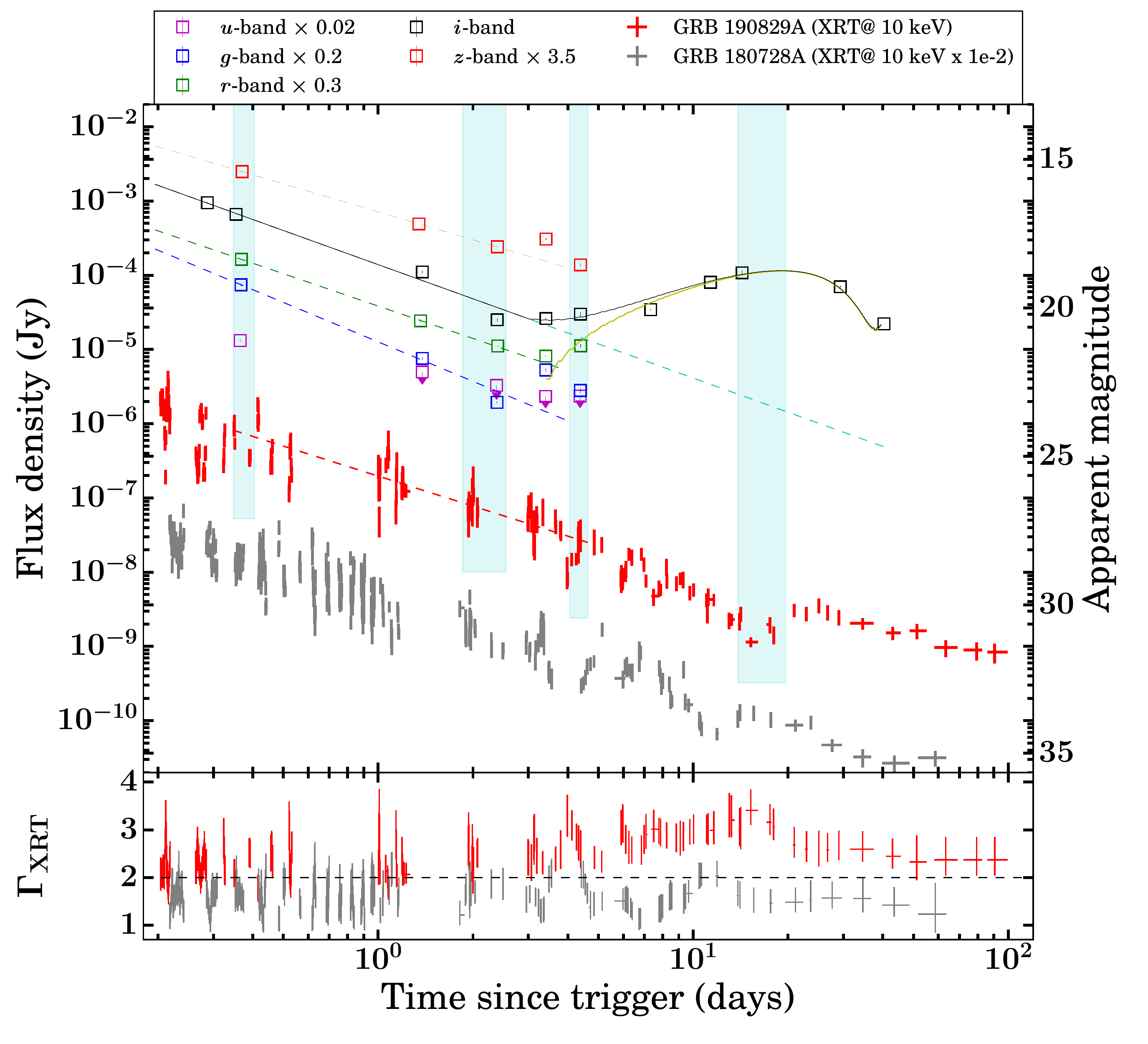}
  \caption{10.4m GTC multi-band optical light curves (in flux density units) of GRB 190829A/SN 2019oyw between 0.32 to 40.3 days post burst. The data is corrected for the galactic and host extinctions, as discussed above. 
  The \textit{i}-band light curve of SN 2019oyw seems to peak around 20 days after the burst which interestingly appears to match with the late-time bump in the 10 keV {\it Swift}/XRT light-curve (in red). 
  The shaded vertical bars (in cyan) show the four epochs used to create the spectral energy distributions of the GRB 190829A afterglow.
   For comparisons, the X-ray light curve (at 10 keV) of GRB 180728A (in gray) is also plotted, it seems to have similar temporal
   features (including temporal decay indices, light curve variability, late-time bump). The bottom panel shows the late time ($\sim$ 4-20
   days post-burst) evolution of the XRT photon indices ($\Gamma_{\rm XRT}$) of the two GRBs discussed. The horizontal black dashed line shows  $\Gamma_{\rm XRT}$ equal to 2.}
  \label{light_curve}
\end{figure}

 \begin{figure}
 \centering
  \includegraphics[scale=0.33]{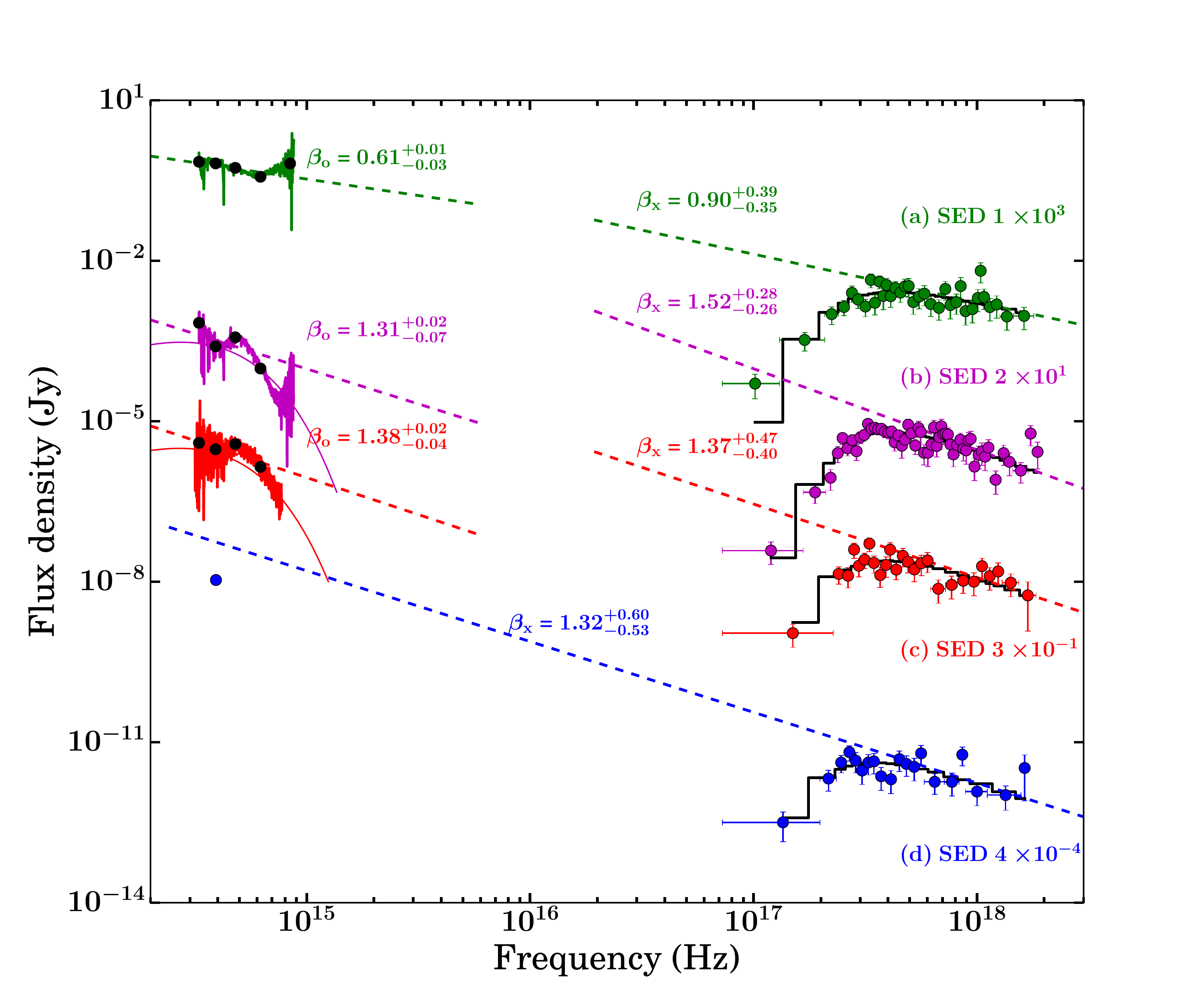}
  \caption{Spectral energy distributions of the GRB 190829A afterglow using the optical-XRT data. (a) SED at $\sim$ 0.37 days (green), (b) SED at $\sim$ 2.2 days (magenta), (c) SED at $\sim$ 4.34 days (red), and (d) SED at $\sim$ 14.3 days (close to the $i$-band peak) showing the absence of thermal emission at XRT frequencies. In (b) and (c) optical SEDs, fitted BB contributions are also plotted in respective colours as described in section \ref{opticalsp}. The derived values of the spectral indices are tabulated in Table \ref{tab:SEDs}.}
  \label{SEDs}
\end{figure}

\section{10.4m GTC Photometric Observations of GRB 190829A/SN 2019oyw}
\label{Photometric}
Multi-band optical photometry obtained for GRB 190829A/SN 2019oyw were measured using aperture photometry through standard procedures after image subtraction. A rotated template for the elliptical galaxy was used to estimate the contribution from the host assuming a symmetric light distribution. 
The photometry was calibrated against a number of stars in the field of view of the GRB/SN from the SDSS catalogue~\citep{2015ApJS..219...12A}. These calibration stars were chosen so that they are isolated, not saturated and are in a constant background. The derived AB magnitudes of the afterglow/SN 2019oyw in SDSS ($ugriz$) filters are tabulated in Table \ref{tab:photometry}.

The optical photometric observations of GRB 190829A/SN 2019oyw span from 0.29 to 40.33 days after the trigger and together with the {\tt XRT} lightcurves (at 10 keV) in the observer-frame are shown in Fig. \ref{light_curve}. The observed magnitudes have been corrected for Galactic as well as the host galaxy extinction using E(B-V) = 0.049 mag and 1.04 mag, respectively. The host-galaxy extinction is adopted from the best fitted model of \cite{2020ApJ...898...42C} obtained for the Small Magellanic Cloud (SMC) extinction law. The plotted \textit{ugriz} lightcurves of GRB 190829A/SN 2019oyw are contributions of the AG and the associated SN, whereas the constant flux contribution from the host galaxy has already been removed using the template subtraction technique. Late time \textit{i}-band (in black) data (up to $\sim$ 40 days after the burst) of GRB 190829A/SN 2019oyw clearly shows the expected SN signature in terms of the late-time bump peaking around $\sim$20 days. To extract the afterglow contribution from the \textit{i}-band data, we fitted the first four epochs (up to $\sim$3 days post-burst) with a single power-law model having an average temporal flux decay index of $\alpha_{\rm opt}$ $\sim$1.53 $\pm$ 0.13 using $griz$ data. 

\subsection{Light curve evolution}
Using these temporal decay indices, we extrapolated the AG contribution up to $\sim$ 41 days (see Fig.\ref{light_curve}, cyan dashed line) and subtracted it from the \textit{i}-band light-curve to get the resultant light curve of the associated SN (in lime green). The associated SN 2019oyw appears to emerge at very early phases (after $\sim$ 3 days) and seems to reach peak magnitude at $\sim$ 20 days. The X-ray light curve (at 10 keV) shows a late time bump at $\sim$ 20 days, contemporaneous with the SN bump, see Fig. \ref{light_curve} (in red), however, we do not see any black-body evolution at the four epochs within the joint Spectral Energy Distributions (SEDs, see the Fig. \ref{SEDs}). Interestingly, such late time XRT bumps at 10 keV are also visible in the cases of GRB 171205A \citep{2019Natur.565..324I} and GRB 180728A \citep{2019ApJ...874...39W, 2020ApJ...893..148R}. Though, based on observations, the nature of possible progenitors and powering mechanisms for these nearby rare events are constrained in terms of the plausible models such as ``Collapsar'' \citep{1993ApJ...405..273W} and binary-driven hypernova model \citep{2001ApJ...555L.117R, 2016ApJ...832..136R}. The detailed understanding of this sub-class of nearby bursts i.e. GRB 171205A/SN 2017iuk, GRB 180728A/SN 2018fip and GRB 190829A/SN 2019oyw having noticeable late-time X-ray bumps (at 10 keV) along with evolution of XRT photon indices ($\Gamma_{\rm XRT}$) is subject of a separate detailed investigation.

\subsubsection{SED evolution of GRB 190829A}
As described in Fig. \ref{SEDs}, near-simultaneous optical and X-ray SEDs (shown in cyan shaded-bands) were constructed at four different epochs covering the afterglow decay phase and the peak of SN 2019oyw. The SEDs at XRT frequencies were collected from the {\it Swift}-XRT page\footnote{\url{https://www.swift.ac.uk/xrt_spectra/}}
and modeled using XSPEC to determine the spectral indices. The 10.4m GTC extinction corrected SEDs were fitted using a single power-law model as discussed above. The X-ray temporal decay index using data taken at 10 keV was found to be $\alpha_{\rm x-ray} \sim$ 1.34$^{+0.06}_{-0.06}$ between $\sim$ 3 $\times$ $10^{4}$ to $\sim$ 4 $\times$ $10^{5}$ s. This temporal index along with those estimated at optical frequencies $\alpha_{\rm opt}$ were used to study the evolution of the SEDs. The details of the four epochs of SEDs and their corresponding indices for the different segments of optical and X-ray data (SED 1- SED 4) are listed in Table \ref{tab:SEDs}. We used $\alpha_{\rm opt}-\beta_{\rm opt}$, $\alpha_{\rm x-ray}-\beta_{\rm x-ray}$ closure relations \citep{1998ApJ...497L..17S, 2018ApJ...866..162G} to constrain the model and location of the cooling-break frequency ($\nu_c$). Considering adiabatic cooling without energy injection from the central engine and a slow cooling case for an interstellar matter (ISM)-like environment as suggested by \cite{2020ApJ...898...42C} and \cite{2020arXiv200311252F}, there are three possible scenarios from optical to X-ray frequencies \citep{2018ApJ...866..162G}. \\
\newline
(i) $\nu_{\rm c}$ $<$  $\nu_{\rm opt}$  $<$ $\nu_{\rm x-ray}$ 
; in this case
\begin{center}
$\alpha_{\rm opt} = \alpha_{\rm x-ray} = \frac{3p-2}{4}$; $\beta_{\rm opt} = \beta_{\rm x-ray} = \frac{p}{2}$
\end{center}
(ii) $\nu_{\rm opt}$ $<$ $\nu_{\rm c}$ $<$ $\nu_{\rm x-ray}$; in this case
\begin{center}
$\alpha_{\rm opt} = \frac{3(p-1)}{4}$; $\alpha_{\rm x-ray} = \frac{3p-2}{4}$; $\beta_{\rm opt} = \frac{p-1}{2}$; $\beta_{\rm x-ray} = \frac{p}{2}$
\end{center}
(iii) $\nu_{\rm opt}$ $<$ $\nu_{\rm x-ray}$ $<$ $\nu_{\rm c}$; in this case
\begin{center}
$\alpha_{\rm opt} = \alpha_{\rm x-ray} = \frac{3(p-1)}{4}$; $\beta_{\rm opt} = \beta_{\rm x-ray} = \frac{p-1}{2}$
\end{center}

We calculated the electron distribution index ($p$) for each scenario mentioned above using the calculated value of $\alpha_{\rm opt,x-ray}$ and $\beta_{\rm opt,x-ray}$. We found that for SED 1, the afterglow could be described with the $\nu_{\rm opt}$ $<$ $\nu_{\rm c}$ $<$ $\nu_{\rm x-ray}$ spectral regime. Later on, for SED 2, SED 3 and SED 4, $\nu_{\rm c}$ seems to have crossed the optical/X-ray band, and we are now in the $\nu_{\rm c}$ $<$  $\nu_{\rm opt}$  $<$ $\nu_{\rm x-ray}$ spectral regime.
\begin{table}
\tiny
\centering
\caption{The best fit optical and X-ray spectral indices at different epoch SEDs and their best describe spectral regime. p is the mean value of the electron distribution indices calculated from observed value of $\alpha_{\rm opt}$/$\alpha_{\rm x-ray}$ and $\beta_{\rm opt}$/$\beta_{\rm x-ray}$ of best describe spectral regime.}
\label{tab:SEDs}

\begin{tabular}{|c|c|c|c|c|}
\hline
\textbf{SED} & \textbf{Time interval (s)} & \textbf{$\bf \beta_{\rm \bf opt}$} & \textbf{$\bf \beta_{\rm \bf x-ray}$} & \textbf{\begin{tabular}[c]{@{}c@{}}p\\ (Spectral regime)\end{tabular}} \\ \hline
1 & 3-3.5 $\times$ $10^{4}$ & 0.61$^{+0.01}_{-0.03}$ & 0.90$^{+0.39}_{-0.35}$ & \begin{tabular}[c]{@{}c@{}}2.33 $\pm$ 0.38\\ ($\nu_{\rm opt}$ $<$ $\nu_{\rm c}$ $<$ $\nu_{\rm x-ray}$)\end{tabular} \\ \hline
2 & 1.6-2.2 $\times$ $10^{5}$ & 1.31$^{+0.02}_{-0.07}$ & 1.52$^{+0.28}_{-0.26}$ & \begin{tabular}[c]{@{}c@{}}2.65 $\pm$ 0.23\\ ($\nu_{\rm c}$ $<$  $\nu_{\rm opt}$  $<$ $\nu_{\rm x-ray}$)\end{tabular} \\ \hline
3 & 3.5-4.0 $\times$ $10^{5}$ & 1.38$^{+0.02}_{-0.04}$ & 1.37$^{+0.47}_{-0.40}$ & \begin{tabular}[c]{@{}c@{}}2.61 $\pm$ 0.14\\ ($\nu_{\rm c}$ $<$  $\nu_{\rm opt}$  $<$ $\nu_{\rm x-ray}$)\end{tabular} \\ \hline
4 & 1.2-1.7 $\times$ $10^{6}$ & - & 1.32 $^{+0.60}_{-0.53}$ & \begin{tabular}[c]{@{}c@{}}2.54 $\pm$ 0.09\\ ($\nu_{\rm c}$ $<$  $\nu_{\rm opt}$  $<$ $\nu_{\rm x-ray}$)\end{tabular} \\ \hline

\end{tabular}
\end{table}

\subsection{SN 2019oyw properties and comparisons}
\begin{figure}
 \centering
  \includegraphics[scale=0.75]{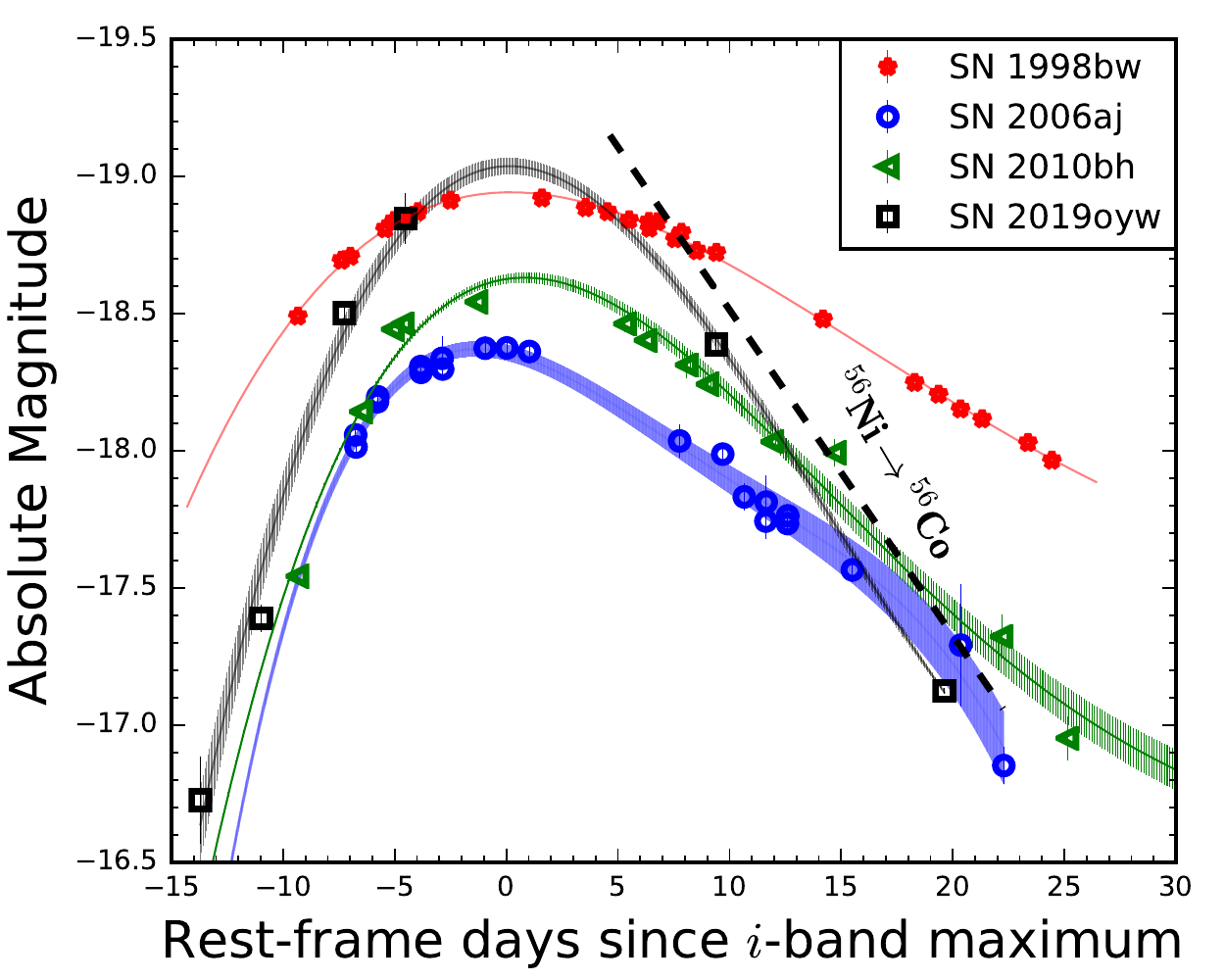}
  \caption{The \textit{i}-band light curve of SN 2019oyw is presented along with three other well studied GRB-SN events: SN 1998bw, SN 2006aj, and SN 2010bh. Lightcurves for the given SNe have been de-reddened (Galactic + Host absorption) and also magnitude values and phase have been shifted to the rest-frame. The constrained value of the peak brightness of SN 2019oyw is close to that estimated in the case of SN 1998bw. Whereas, SN 2019oyw exhibits a post-peak decay rate close to that expected by $^{56}Ni$ --> $^{56}Co$ decay.}
  \label{SN_LC}
\end{figure}

The extracted lightcurve of SN 2019oyw in absolute magnitudes (in black) is plotted in the rest-frame, see Fig. \ref{SN_LC}. The absolute magnitudes are calculated from de-reddened apparent magnitudes and also corrected for cosmological expansion \citep{Hogg2002} to get the rest-frame magnitudes, as described in \cite{Kumar2020}. The \textit{i}-band light-curve evolution of SN 2019oyw (see Fig.\ref{SN_LC}, black solid line) is compared with other well studied low-redshift GRB-SNe such as SN 1998bw \citep[in red;][]{Galama1998}, SN 2006aj \citep[in blue;][]{Bianco2014}, and SN 2010bh \citep[in green;][]{Cano2011} after correcting for Galactic as well as host galaxy extinction using values taken from \cite{Cano2013} and references therein.

Lightcurves of the four SNe connected to GRBs (see Fig. \ref{SN_LC}) are fitted with low-order polynomials to estimate the peak absolute magnitudes ($\rm M_i$) and the times taken by the SN to rise and fall by 1 mag from the peak ($\rm t^{\rm \Delta 1mag}_{\rm rise}$ and $\rm t^{\rm \Delta 1mag}_{\rm fall}$, respectively). Based on our analysis, SN 2019oyw appears to have $\rm M_i$ $=-$19.04 $\pm$ 0.01 mag, brighter than SN 2006aj ($\rm M_i$ $=-$18.36 $\pm$ 0.13 mag) and SN 2010bh ($\rm M_i$ $=-$18.58 $\pm$ 0.08 mag), and close to the peak absolute magnitude of SN 1998bw ($\rm M_i$ $\sim-$18.95 mag). The calculated value of $\rm t^{\rm \Delta 1mag}_{\rm fall}$ for SN 2019oyw is found to be 13.32 $\pm$ 0.04 days, which is lower in comparison to those estimated for other GRB-SNe: SN 1998bw ($\sim$ 24.8 days), SN 2006aj ($\sim$20.3 days), and SN 2010bh ($\sim$ 17.1 days). This indicates a comparatively steeper post-peak decay rate of SN 2019oyw as can also be inferred from the faster spectral evolution in section~\ref{opticalsp}. On the other hand, the value of $\rm t^{\rm \Delta 1mag}_{\rm rise}$ for SN 2019oyw (9.67 $\pm$ 0.02 days) is similar to SN 2006aj ($\sim$ 8.80 days) and SN 2010bh ($\sim$ 9.0 days), whereas it is lower in comparison to that observed in the case of SN 1998bw ($\sim$ 13.3 days). We also estimated the value of the ejected nickel mass ($\rm M_{Ni}$) as $0.5{\pm}0.1\,{\rm M}_{\odot}$ for GRB 190829A/SN 2019oyw using the relation between $\rm M_{\rm peak}$ and $\rm M_{\rm Ni}$ given by \cite{Lyman2016}. The discussed light curves of the four GRB-SNe have also been compared with the $\rm ^{56}Ni$ --> $\rm ^{56}Co$ theoretical decay curve. SN 2019oyw appears to be consistent with this decay curve (shown with a black dotted line) soon after the peak.

As studied by \cite{Wheeler2015}; see also \cite{Cano2013}, we can estimate the ejecta mass ($\rm M_{\rm ej})$ and kinetic energy ($\rm E_k$) of the SN using the photospheric velocity near the peak and rise time (see equations 1 and 3 of \citealt{Wheeler2015}). SN 2019oyw exhibits a rise time ($\rm t_r$, i.e. the time taken from the GRB detection to the SN peak time in the \textit{i}-band) of 19.19 $\pm$ 0.25 days.  We are unable to estimate the photospheric velocity ($\rm v_{ph}$) for SN 2019oyw due to the absence of late-time spectral coverage. So, we used the average value of photospheric velocity ($\sim20,000$ $\pm$ 2500 km $\rm s^{-1}$) of GRB/XRF-SNe estimated by \cite{Cano2013}. For SN 2019oyw, using $\rm t_r$ and  $\rm v_{\rm ph}$ we constrained $\rm M_{\rm ej}$ 5.67 $\pm$ 0.72 $\rm M_\odot$ and $\rm E_k$ (13.55 $\pm$ 5.08) $\times$ 10$^{51}$ erg. For this analysis, the fiducial optical opacity $\kappa$ = 0.1 $\rm cm^2$ $\rm g^{-1}$ and the fiducial gamma-ray opacity $\kappa_\gamma$ = 0.03 $\rm cm^2$ $\rm g^{-1}$ are adopted as suggested by \cite{Wheeler2015}. The estimated values of $\rm M_{Ni}$, $\rm M_{ej}$ and $\rm E_k$ for SN 2019oyw are well in agreement with those estimated for SN 1998bw by \cite{2001ApJ...550..991N}. For SN 2019oyw, the $\rm E_k$/$\rm M_{\rm ej}$ is also well consistent with the values estimated by \cite{Lyman2016}, \cite{Cano2017}, and \cite{Pandey2020}.

\section{Results and Discussion}
 \label{results}
The prompt emission light curve of GRB 190829A, with the two emission episodes separated by a quiescent gap, is found to be very 
much similar to that exhibited by another nearby GRB 180728A consisting of a fainter precursor followed by the brighter main pulse. Time-averaged and time-resolved spectral analysis of the double-episode prompt emission of {\it Swift}/BAT and {\it Fermi}/GBM data from GRB 190829A and GRB 180728A exhibit diverse $\rm E_{\rm p}$ and $\alpha$ evolution posing a challenge to the proposed progenitor models. Particularly, our
spectral analysis indicates that in the case of GRB 190829A, the low-energy spectral index ($\alpha$) seems to overshoot the synchrotron limits in later bins posing problem for synchrotron origin whereas 
in the case of GRB 180728A the evolution of $\alpha$ remains within the synchrotron limit.\\

10.4m GTC spectroscopy (0.32 to 4.09 days post-burst, in the rest-frame) and the redshift determination of the VHE-detected GRB 190829A are also presented in this article. Our spectrum taken as early as 0.32 days shows a featureless power-law behaviour as expected from afterglows whereas the spectrum taken at a later epoch (4.09 days post-burst) shows the type Ic-BL broad absorption features (Si II and Ca II NIR lines) indicative of higher velocities as reported by \cite{2019GCN.25677....1D}. Thermal evolution of the spectra at three later epochs shows decreasing photospheric temperatures from $\sim$5100 to 4575 K, typical of those observed in other similar SNe\citep{Cano2017}. Spectroscopically, around 4.09 days post-burst, the underlying SN closely resembles GRB 980425/SN 1998bw.\\ 
 
 10.4m GTC $ugriz$ band photometry produced after calibration and host galaxy subtraction were used to construct the light curves  (0.27 to 37.99 rest-frame days after burst) which clearly shows a power-law decay nature as expected in afterglows till 2-3 days post burst.  Using 10.4m GTC multi-band optical data along with the XRT data, we were also able to constrain the evolution
 of $\nu_c$ (between 0.32 to 4.09 days post-burst) and the electron energy index $p$ considering the afterglow follows the model predictions in the case of an ISM-like ambient medium. However, photometrically, apart from showing a typical afterglow decay at early epochs, a deviation from the power-law decay is clearly seen in all filters with a clear signature of a re-brightening peaking around $\sim$ 20 days post-burst in the $i$-band. The peak brightness (M$_i \sim$ -19.04 mag) of SN 2019oyw confirms the SN as being one of the brightest GRB/SNe closely matching SN 1998bw in terms of other estimated parameters such as $\rm M_{\rm Ni}$, $\rm M_{\rm ej}$ and $\rm E_k$. Overall, while the values of $\rm M_{\rm ej}$ and $E_k$ are higher than usual, their ratio is closer to those seen in other type Ibc-BL and GRB/SNe \citep{Lyman2016,Cano2017}. The lower value of $\rm t^{\rm \Delta 1mag}_{\rm fall}$, and the post-peak decay rate are indicative of $\rm ^{56}Ni$ as a possible powering source for SN 2019oyw.\\
 
 It is also notable that \cite{2019ApJ...874...39W} have attempted explaining the nature of the prompt emission of GRB 180728A in terms of the type II binary-driven hypernova (BdHN II) model for the observed underlying supernova as an alternative to the ``collapsar'' model~\citep{1993ApJ...405..273W, 2012grb..book..169H}. In the near future, with more such observed events, it would be very interesting to know if such nearby GRBs having two emission episodes in their prompt emission phase have underlying SN features.

 \begin{acknowledgements}
 Partly based on observations made with the Gran Telescopio Canarias (GTC), installed at the Spanish Observatorio del Roque de los Muchachos of the Instituto de Astrofísica de Canarias, in the island of La Palma. Y.-D-H., A.J.C.-T., I.A. and D.A.K. acknowledge financial support from the State Agency for Research of the Spanish MCIU through the "Center of Excellence Severo Ochoa" award to the Instituto de Astrofísica de Andalucía (SEV-2017-0709). RG, SBP, and AA acknowledge BRICS grant DST/IMRCD/BRICS/PilotCall1/ProFCheap/2017(G) for part of the work. VAF was  supported  by  RFBR 18-29-21030  grant. SBP also acknowledge DST/INT/JSPS/P/281/2018 for the present work.  BBZ acknowledges the supported by the Fundamental Research Funds for the Central Universities (14380035). This work is supported by National Key Research and Development Programs of China (2018YFA0404204), the National Natural Science Foundation of China (Grant Nos. 11833003) and the Program for Innovative Talents, Entrepreneur in Jiangsu. D.A.K. acknowledges support from Spanish research project RTI2018-098104-J-I00 (GRBPhot).
 \end{acknowledgements}
 
\bibliographystyle{aa} 

\end{document}